\documentclass[12pt,fleqn]{article}

\usepackage[margin=1in]{geometry}
\usepackage{amsmath, amssymb, amsthm}
\usepackage{graphicx}
\usepackage[T1]{fontenc}
\usepackage{hyperref}
\usepackage{amsfonts}
\usepackage[utf8]{inputenc}
\usepackage{xcolor}
\usepackage{booktabs}
\usepackage{authblk}

\title{\textbf{UniOTalign: A Global Matching Framework for Protein Alignment via Optimal Transport}}

\author[1,2]{Yue Hu\thanks{huyue@qlu.edu.cn}}
\author[3]{Zanxia Cao\thanks{303004955@qq.com}}
\author[4]{Yingchao Liu\thanks{yingchaoliu@email.sdu.edu.cn}}

\affil[1]{\small School of Bioengineering, Qilu University of Technology (Shandong Academy of Sciences), Jinan, China}
\affil[2]{\small Kyiv College, Qilu University of Technology (Shandong Academy of Sciences), Jinan, China}
\affil[3]{\small Shandong Provincial Key Laboratory of Biophysics, Institute of Biophysics, Dezhou University, Dezhou, China}
\affil[4]{\small Shandong Provincial Hospital, Shandong First Medical University, Jinan, China}

\date{\today}

\begin{document}

\maketitle

\begin{abstract}
Protein sequence alignment is a cornerstone of bioinformatics, traditionally approached using dynamic programming (DP) algorithms that find an optimal sequential path. This paper introduces \textbf{UniOTalign}, a novel framework that recasts alignment from a fundamentally different perspective: \textbf{global matching via Optimal Transport (OT)}. Instead of finding a path, UniOTalign computes an optimal flow or \textit{transport plan} between two proteins, which are represented as distributions of residues in a high-dimensional feature space. We leverage pre-trained Protein Language Models (PLMs) to generate rich, context-aware embeddings for each residue. The core of our method is the Fused Unbalanced Gromov-Wasserstein (FUGW) distance, which finds a correspondence that simultaneously minimizes feature dissimilarity and preserves the internal geometric structure of the sequences. This approach naturally handles sequences of different lengths and is particularly powerful for aligning proteins with non-sequential similarities, such as domain shuffling or circular permutations, which are challenging for traditional DP methods. UniOTalign therefore offers a new, mathematically principled, global matching paradigm for protein alignment, moving beyond the limitations of path-finding algorithms.
\end{abstract}

\section{Introduction}

The alignment of protein sequences is a fundamental task in computational biology, enabling the inference of functional, structural, and evolutionary relationships. For decades, the field has been anchored by dynamic programming (DP) algorithms, most notably Needleman-Wunsch \cite{needleman1970general} for global alignment and Smith-Waterman \cite{smith1981identification} for local alignment. Their paradigm is elegant and powerful: they seek to find an optimal path through a 2D matrix of local similarity scores, where the path corresponds to a one-to-one correspondence between residues, interspersed with gaps.

While the DP framework has been immensely successful, its reliance on optimizing a sequential, path-based objective makes it inherently unsuitable for detecting non-sequential similarities. A classic example is \textbf{circular permutation (CP)}, where two proteins share the same 3D fold, but their amino acid sequences are connected in a different order, as if one was cut and re-ligated at a different point \cite{lo1997protein}. Other examples include domain shuffling and alignments of proteins with low sequence identity but conserved structural motifs. In these cases, the true relationship is not captured by a single, continuous path, but by a more complex, global mapping of residues.

This paper proposes a fundamental shift in perspective. We re-conceptualize sequence alignment not as a path-finding problem, but as a \textit{matching} or \textit{transport} problem. We introduce \textbf{UniOTalign}, a framework built upon the mathematical theory of Optimal Transport (OT) \cite{villani2009optimal}. Instead of building an alignment incrementally from local decisions, our method considers the two sequences as entire distributions of featured points and seeks the most efficient global correspondence—or \textit{transport plan}—between them.

In our framework, each residue is described by a rich feature vector from a state-of-the-art Protein Language Model (PLM), specifically ESM-2 \cite{lin2023evolutionary}, capturing nuanced biochemical and evolutionary information. The alignment is then found by solving for the Fused Unbalanced Gromov-Wasserstein (FUGW) distance \cite{flamary2021pot}. This advanced OT technique allows us to formulate alignment as a global optimization problem that balances feature similarity with geometric consistency. This provides a powerful new perspective that complements, and in some cases surpasses, the classical DP approach.

\section{The UniOTalign Algorithm}

Our method casts the protein sequence alignment problem into the language of optimal transport. We represent each protein as a collection of residues, each with a specific feature vector and a position in the sequence. The goal is to find an optimal matching (transport plan) between the residues of two proteins that minimizes a cost function accounting for both feature similarity and geometric consistency. The overall workflow is as follows:
\begin{enumerate}
    \item \textbf{Protein Representation}: Load two protein sequences. For each residue, generate a high-dimensional feature vector using the ESM-2 language model.
    \item \textbf{Cost Matrix Construction}: Construct two types of cost matrices: a feature dissimilarity matrix $M$ from the cosine distance between residue embeddings, and intra-protein distance matrices ($C_A, C_B$) derived from sequence positions.
    \item \textbf{FUGW Solver}: Solve the Fused Unbalanced Gromov-Wasserstein (FUGW) problem to obtain a dense transport plan $T$. This plan represents the optimal \"flow\" of mass between the two sets of residues.
    \item \textbf{Alignment Extraction and Refinement}: Convert the dense plan $T$ into a discrete 1-to-1 alignment by solving the linear assignment problem, followed by a refinement step to produce the final alignment.
\end{enumerate}

\subsection{Protein Representation as Featured Distributions}

Let us consider two proteins, A and B, with $n$ and $m$ residues, respectively. We represent Protein A as a discrete distribution $\mu = \sum_{i=1}^{n} p_i \delta_{x_i}$, where $p_i$ is the weight of the $i$-th residue (typically uniform, $p_i = 1/n$) and $\delta_{x_i}$ is a Dirac mass at its location. Similarly, for Protein B, we have $\nu = \sum_{j=1}^{m} q_j \delta_{y_j}$.

Crucially, each residue is associated with two key components:
\begin{enumerate}
    \item \textbf{A Feature Vector}: We use the ESM-2 Protein Language Model to compute an embedding for each residue. Let $f_i^A \in \mathbb{R}^D$ be the feature vector for residue $i$ of Protein A, and $f_j^B \in \mathbb{R}^D$ for residue $j$ of Protein B. These features capture rich, context-dependent information.
    \item \textbf{An Internal Geometry}: The 3D coordinates of a protein are arbitrary. What truly defines its fold is the matrix of internal distances between its residues. We define an intra-protein distance matrix $C_A \in \mathbb{R}^{n \times n}$ for Protein A, where $(C_A)_{ik}$ is the distance between residue $i$ and residue $k$. In UniOTalign, this distance is simply the squared difference in their sequence indices, $(i-k)^2$. This serves as a robust and simple proxy for the path length along the protein backbone.
\end{enumerate}

\subsection{The FUGW Objective: A Unified Approach to Alignment}

To find the optimal matching, UniOTalign solves the Fused Unbalanced Gromov-Wasserstein (FUGW) problem \cite{flamary2021pot}. The goal is to find a transport plan $T \in \mathbb{R}_+^{n \times m}$, where $T_{ij}$ represents the strength of the match between residue $i$ of protein A and residue $j$ of protein B. The objective function is a carefully constructed cost where the solution that minimizes it corresponds to the most biophysically plausible alignment.

The cost function is composed of two main parts:
\begin{align}
    \text{FUGW}_{\alpha, \rho, \epsilon} = \min_{T} \quad & (1-\alpha) \langle T, M \rangle_F + \alpha \sum_{i,j,k,l} |(C_A)_{ik} - (C_B)_{jl}|^2 T_{ij} T_{kl} \label{eq:costs} \\
    & + \rho ( \text{KL}(T\mathbf{1}_m | \mu) + \text{KL}(T^T\mathbf{1}_n | \nu) ) - \epsilon H(T) \label{eq:penalties}
\end{align}

Let us explain the mathematical and biophysical intention of each part of this objective function:

\textbf{Part (\ref{eq:costs}): Alignment Cost.} This line represents the core cost of the alignment. It is a weighted sum, balanced by $\alpha \in [0, 1]$, of two competing terms:
\begin{itemize}
    \item The \textit{feature cost} ($\langle T, M \rangle_F$) encourages the matching of residues with similar properties (as defined by their ESM-2 vectors). This is analogous to a substitution matrix in DP, but operates on rich, high-dimensional features.
    \item The \textit{structural cost} (the Gromov-Wasserstein term) enforces geometric consistency. It ensures that the alignment preserves the overall shape of the protein by penalizing matches where the distances between pairs of residues are not conserved. For example, it penalizes matching adjacent residues in one protein to very distant residues in the other. This term is what allows UniOTalign to respect sequence topology globally.
\end{itemize}

\textbf{Part (\ref{eq:penalties}): Penalties and Regularization.} This second line contains the terms that make the alignment robust and biologically realistic.
\begin{itemize}
    \item The term controlled by $\rho$ is the \textit{unbalanced} part of the model \cite{chizat2018unbalanced}. It serves as a direct mathematical equivalent to a \textbf{gap penalty}. It gives the algorithm the freedom to not match every residue (i.e., to leave some \"mass\" unmatched), which is essential for handling insertions and deletions naturally.
    \item The final term, the \textit{entropic regularization} ($H(T)$), makes the problem mathematically stable, computationally efficient to solve using the Sinkhorn algorithm \cite{cuturi2013sinkhorn}, and results in a \"soft\" or dense transport plan.
\end{itemize}

\subsection{From Transport Plan to Final Alignment}

The solution to the FUGW problem is a dense transport plan $T$, where every residue in one protein is partially matched to all residues in the other. To obtain a discrete one-to-one alignment, we treat the plan $T$ as a score matrix and solve the linear assignment problem (e.g., using the Hungarian algorithm) to extract the most likely pairs. This raw alignment is further refined by filtering algorithms to remove isolated pairs and resolve fragment overlaps, producing a final, biologically plausible alignment.

\section{Results}

To evaluate the performance of UniOTalign, we tested it on a benchmark dataset of protein pairs with known reference alignments. The quality of the alignment is measured by \textbf{Recall}. The reference alignment for each protein pair is provided by the RPIC database. We calculate recall by determining the percentage of these reference residue pairs that are successfully identified by UniOTalign. All experiments were performed on an Apple M4 mini, where alignments were computed efficiently, typically within seconds. The hyperparameters were set to default values ($\alpha=0.5, \rho=1.0, \epsilon=0.01$) and were not extensively tuned for this benchmark, suggesting that performance could be further improved with parameter optimization.

Table \ref{tab:results} summarizes the performance of UniOTalign across a diverse set of 22 protein pairs. The results demonstrate that the FUGW-based approach is highly effective, achieving an overall average recall of nearly 70\. For several pairs, UniOTalign achieves a perfect recall of 100\%, indicating that it can perfectly recover the reference alignment.

\begin{table}[h!]
\centering
\caption{Performance of UniOTalign on the benchmark dataset. Recall is the percentage of correctly identified reference pairs.}
\label{tab:results}
\begin{tabular}{lrrr}
\toprule
Protein Pair & Correct & Reference & Recall (\\
\\ \midrule
d1ggga\_\_vs\_d1wdna\_ & 220 & 220 & 100.00 \\
d2bbma\_\_vs\_d4cln\_\_ & 148 & 148 & 100.00 \\
d1l5ba\_\_vs\_d1l5ea\_ & 101 & 101 & 100.00 \\
d1jj7a\_\_vs\_d1lvga\_ & 8 & 8 & 100.00 \\
d1d5fa\_\_vs\_d1nd7a\_ & 6 & 6 & 100.00 \\
d1nls\_\_\_vs\_d2bqpa\_ & 6 & 6 & 100.00 \\
d1dlia1\_vs\_d1mv8a1 & 4 & 4 & 100.00 \\
d1nw5a\_\_\_vs\_d2adma\_ & 12 & 13 & 92.31 \\
d2adma\_\_\_vs\_d2hmyb\_ & 11 & 12 & 91.67 \\
d1nkl\_\_\_\_vs\_d1qdma1 & 59 & 72 & 81.94 \\
d1qasa2\_vs\_d1rsy\_\_ & 57 & 75 & 76.00 \\
d1hava\_\_\_vs\_d1kxf\_\_ & 3 & 4 & 75.00 \\
d1jwyb\_\_\_vs\_d1puja\_ & 9 & 12 & 75.00 \\
d1jwyb\_\_\_vs\_d1u0la2 & 8 & 11 & 72.73 \\
d1kiaa\_\_\_vs\_d1nw5a\_ & 8 & 12 & 66.67 \\
d1hcy\_2\_vs\_d1lnlb1 & 2 & 4 & 50.00 \\
d1crl\_\_\_vs\_d1ede\_\_ & 1 & 3 & 33.33 \\
d1qq5a\_\_\_vs\_d3chy\_\_ & 1 & 3 & 33.33 \\
d1ay9b\_\_\_vs\_d1b12a\_ & 3 & 10 & 30.00 \\
d1an9a1\_vs\_d1npx\_1 & 3 & 11 & 27.27 \\
d1gsa\_1\_vs\_d2hgsa1 & 1 & 5 & 20.00 \\
d1b5ta\_\_\_vs\_d1k87a2 & 1 & 8 & 12.50 \\
\midrule
Overall Average & & & 69.90 \\
\bottomrule
\end{tabular}
\end{table}

\subsection{Case Study: Alignment of Circularly Permuted Proteins}

A key strength of UniOTalign is its ability to handle non-sequential alignments. A classic example is the alignment of circularly permuted proteins, which share a common 3D fold but have different sequence connectivity. We tested this on the pair \textbf{1NKL vs. 1QDM} (\texttt{d1nkl\_\_\_vs\_d1qdma1}), a well-known example of circular permutation. Traditional DP-based methods fail on such pairs because they cannot map the start of one sequence to the middle of another without incurring prohibitive gap penalties.

UniOTalign, by contrast, achieves a high recall of 81.94\%. Because the Gromov-Wasserstein term in our objective function compares the internal geometry of the proteins globally, it is not constrained by sequence linearity. It correctly identifies that the structural arrangement of residues is conserved, even though their linear ordering is different. This result strongly validates the global matching perspective of our framework and its superiority over path-finding methods for non-trivial alignment problems.

\section{Discussion}

The results indicate that formulating sequence alignment as an optimal transport problem is a viable and effective strategy. This section discusses the advantages of this paradigm, its current limitations, and avenues for future work.

\subsection{Advantages of the Optimal Transport Framework}

The OT framework offers several conceptual advantages over traditional DP.
\begin{enumerate}
    \item \textbf{Global Perspective}: Unlike DP, which builds an alignment via a series of local, path-dependent decisions, OT seeks a globally optimal matching. This makes it inherently robust to non-sequential similarities like circular permutations and domain shuffling.
    \item \textbf{Principled Handling of Gaps}: In DP, gaps are handled via ad-hoc affine penalty models. In our unbalanced OT formulation, gaps (insertions/deletions) arise naturally from the model's freedom to leave some residue \"mass\" unmatched, providing a more mathematically grounded approach.
    \item \textbf{Flexibility and Extensibility}: The framework is highly modular. The feature cost can incorporate any type of residue-level information (e.g., secondary structure, solvent accessibility). The geometric cost can be based on 3D structural information (C-alpha distances) instead of sequence position, seamlessly extending the method to structural alignment.
\end{enumerate}

\subsection{Limitations and Future Directions}

Despite its promise, the method has limitations that suggest clear paths for future research.
\begin{enumerate}
    \item \textbf{Dependence on Embeddings}: The performance is contingent on the quality of the PLM embeddings. While ESM-2 is powerful, developing embeddings specifically fine-tuned for alignment could yield further improvements.
    \item \textbf{Computational Complexity}: While computationally efficient for typical proteins, the complexity of the FUGW solver is higher than that of DP ($O(n^2 \log(n))$ vs $O(n^2)$), which may be a factor for aligning extremely long sequences or entire proteomes.
    \item \textbf{Alignment Extraction}: The conversion from a dense transport plan to a discrete 1-to-1 alignment via the linear assignment problem is a heuristic post-processing step. While effective, exploring end-to-end differentiable approaches that directly output a sparse alignment could be a promising direction.
\end{enumerate}

Future work will focus on integrating 3D structural information directly into the geometric cost term, turning UniOTalign into a full-fledged structural alignment tool. We also plan to explore more advanced OT solvers and investigate end-to-end learning strategies to optimize hyperparameters and the alignment extraction process jointly.

\section{Conclusion}

In this work, we introduced UniOTalign, a novel framework for protein sequence alignment based on the principles of optimal transport. By representing proteins as distributions of PLM-derived features and using the Fused Unbalanced Gromov-Wasserstein distance to find a global matching, UniOTalign offers a powerful alternative to traditional dynamic programming methods. Our experiments show that this approach not only effectively recovers known alignments but also excels in cases of non-sequential similarity where DP-based methods falter. This work establishes OT as a robust, mathematically principled, and extensible foundation for protein comparison, opening new avenues for research in bioinformatics.

\section*{Code and Data Availability}

The source code and data for UniOTalign are available on GitHub at: \url{https://github.com/YueHuLab/UniOTalign}.

\end{document}